# Destructive Creation, Creative Destruction, and the Paradox of Innovation Science

Likun Cao[a], Ziwen Chen[b], James Evans[a,c*]

[a]Sociology Department & Knowledge Lab, University of Chicago

[b]Graduate School of Business, Stanford University

[c]Santa Fe Institute

**Funding Information**

Air Force Office of Scientific Research (AFOSR) #FA9550-19-1-0354; FA9550-15-1-0162.

**Correspondence**

James Evans, 5500 S. University Ave. #210, Chicago, IL 60637. Email: jevans@uchicago.edu

**Acknowledgements**

We thank Amanda Sharkey, Chris Esposito and anonymous reviewers for helpful comments and insights. We thank AFOSR for support in grants associated with prediction and predictability in science and innovation.

**Conflict of Interest**

The authors declare no conflict of interest.

**ORCID**

*Likun Cao* 0000-0001-5234-3855

*Ziwen Chen* 0000-0001-7563-5922

*James Evans* 0000-0001-9838-0707

**Author Biographies:**

**Likun Cao** is a Ph.D. student in sociology at the University of Chicago and a member of Knowledge Lab. Her research focuses on science and technology innovation and its social impacts, specifically on conditions and mechanisms of novelty emergence, and how innovations reshape business competition and social structures. Her work is based on computational methods, including language processing and machine learning, as well as advanced statistical models.

**Ziwen Chen** is a Ph.D. student in Organizational Behavior at Stanford Graduate School of Business. Using computational social science methods (e.g., NLP, online experiments, machine learning), she studies the construction and transmission of culture and innovation through collective action.

**James Evans** is the Max Palevsky Professor of Sociology, Director of Knowledge Lab, and Faculty Director of Computational Social Science at the University of Chicago and External Professor at the Santa Fe Institute. His research uses large-scale data, machine learning and generative models to understand how collectives think and what they know. This involves inquiry into the emergence of ideas, shared patterns of reasoning, and processes of attention, communication, agreement, and certainty, with a special interest in innovation.

**Abstract:** Innovation or the creation and diffusion of new material, social and cultural things in society has been widely studied in sociology and across the social sciences, with investigations sufficiently diverse and dispersed to make them unnavigable. This complexity results from innovation's importance for society, but also the fundamental paradox underlying innovation science: When innovation becomes predictable, it ceases to be an engine of novelty and change. Here we review innovation studies and show that innovations emerge from contexts of discord and disorder, breaches in the structure of prior success, through a process we term *destructive creation*. This often leads to a complementary process of *creative destruction* whereby local structures protect and channel the diffusion of successful innovations, rendering alternatives obsolete. We find that social scientists naturally focus far more on how social and cultural contexts influence material innovations than the converse. We highlight computational tools that open new possibilities for the analysis of novel content and context in interaction, and show how this brings us empirically toward the broader range of possibilities that complex systems and science studies have theorized—and science fiction has imagined— the social, cultural and material structures of innovation conditioning each other's change through cycles of disruption and development.

1. Introduction

Studies of innovation focus on things new to the world, but the topic has a deep social history. Introduced by Marx in the 19th Century (Marx, 2004 [1867]) and revived by Schumpeter in the first half of the 20th (Schumpeter, 1942), theories of innovation emerged with theories of capitalism and urban society and have undergone long term development, inspiring a vast number of empirical investigations (Fagerberg, 2018). In the past decade, more than 10,000 papers on the topic have been published each year according to statistics from the Web of Science. The focus has grown increasingly prominent in many fields, with greater social and business attention to innovation and acclaim for theories of self-driven or endogenous growth in economics. Papers are published routinely in top sociological, economics and general science outlets, but they cover dispersed and disconnected aspects of innovation with distinct vocabularies and have become hard to track, index and evaluate together.

This complexity emerges in part from innovation's widespread importance for society at all scales. New-to-the-world scientific discoveries, technological inventions and their widespread diffusion represent the greatest contribution to modern economic growth and collective prosperity (Jones & Summers, 2020; Oreskes, 2019). Innovations forged or adopted by persons, teams, organizations and societies represent shifts that can dramatically alter and improve their character and efficiency. Moreover, innovation is a fundamentally interactive process involving many social groups over multiple time scales and stages, generating highly differentiated, novel products and processes, from technology and science to cultural artifacts and new social forms (Fleming, 2001; Kremp, 2010; Padgett and Powell, 2012; Uzzi et al., 2013). The multi-stage nature of innovation also results in observations that occur at different points of the innovative process.

A second reason behind disconnects in the literature, however, is a fundamental paradox underlying innovation science. The moment that stable, reproducible patterns of innovation become understood, codified and institutionalized, the subject ceases to be innovation. This is not merely a play on words. When innovation becomes predictable and incorporated within established mechanisms of economy and society, then it ceases to be a source of novelty that defies expectation and drives change. Past innovation results in successively new understandings of "what is new" and how to think about novelty.

Following the same intuition, both Karl Marx and Joseph Schumpeter saw in modern capitalism an engine of innovation, but neither believed it could do so predictably and sustainably (Marx, 2004 [1867]; Schumpeter, 1942). In this way, social studies of innovation have a uniquely insecure epistemological status and discovered patterns of innovation are typically either ephemeral or indicative of innovation's absence. Because contemporary society, culture and technology are well-structured and highly regulated, the genesis of innovations in one domain is typically unleashed when established structures and the forces defending them are broken down. Resulting conflict and chaos anticipate the imagination, invention and adoption of innovation, a process we coin here as *destructive creation*. In the life-course of a single invention-turned-innovation, this precedes a companion process of *creative destruction*, anticipated by Neitzche (Nietzsche, 1956), coined by Sombart and popularized by Schumpeter. In creative destruction, emergent innovations render those before irrelevant as they are disseminated and promoted by social, cultural and technical structures.

In the following, we develop these ideas by studying how they appear across analytical levels and research objects. To evaluate them systematically, we identified innovation-based studies by selecting the subset of scientific articles that mention innovation from prominent journals in all branches of social science. We first identified the most highly cited journals in sociology, management, economics, political science, psychology, communication, and data science, then kept all papers with the keywords "innovation", "invention" and "creation" in their titles, keywords, and abstracts. The time scope for our search was 2011 to 2022. After we developed this collection, we made adjustments by including some classic works and those repeatedly mentioned by authors from other publications, and deleting papers with repeated themes. The details of our search process and adjustment methods are shown in Appendix A. By analyzing these (>100) articles in the context of those that have come before, we generated a conceptual framework that allows us to organize them across two dimensions: by their (1) stage in the process of innovation and (2) the analytical level (macro, meso, micro) at which they are focused. In the final section, we explore the consistency between social science perspectives, develop our theory about the impossibility of a complete innovation science, and suggest how these insights can be complemented by other fields and domains of social and cultural life—especially science studies, complex systems and science fiction.

2. Innovation Genesis as Destructive Creation

The genesis of innovation is often conceptualized as "invention" in prior literature (Padgett & Powell, 2012). How and why do some new ideas, tools, and practices emerge from the old? What kind of social forms breed successful versus failed innovations? And can we accelerate innovation in the age of the internet and global connectivity? Here, we explore conditions for the genesis of innovation by societies, social communities, and individuals. We show that disorder and discord—large and small—are often critical preconditions for innovation genesis. By contrast, the diffusion of innovation replaces broken and competing structures to bring new order.

2.1 The Complex Construction of Innovations

Innovations involve a process of recombination, whereby parts or wholes of prior products and processes are incorporated into something new. Complexity scholars have argued for a deep similarity between technological, social and cultural innovations and the process of biological evolution, whereby sex and horizontal transfer combine genes in new, complex combinations that reproduce or die (Erwin & Krakauer, 2004). Padgett and Powell take this metaphor further and argue that new social and cultural structures are like autocatalytic chemical processes, where cascading reactions form novel self-sustaining cycles (2012) in processes that reflect the biochemical origins of life (Kauffman, 1996).

Cumulative innovations are recursive with each new combination "expanding the adjacent possible"—the space of possible combinations one-step from the current frontier (Kauffman, 1996). This process has been widely observed across Wikipedia, social annotation systems, music, text, and technology (Tria et al., 2014). From this perspective, the system of collective innovation is viewed as a complex system with hierarchical, interrelated subsystems in constant evolution. For example, new technologies are constructed from simpler, existing technologies, which subsequently serve as elements in further combinations, resulting in complex, hierarchical architectures (Arthur, 2007; 2009). This process has been formalized using computer simulation (Arthur & Polak, 2006) and validated empirically (Fink & Teimouri, 2019). In science, this perspective represents new ideas as arising from novel or complex configurations of prior knowledge (Fleming, 2001; Fleming & Sorenson, 2004; Uzzi et al., 2013; Youn et al., 2015). Connecting elements from distant sources in improbable combinations rewires the structure of prior understanding into acknowledged advance (Shi et al., 2015; Shi & Evans, 2019). Interest has grown in the possibility that invention occurs through recombination not only in science and technology, but other facets of social life. For example, protest leaders engaged in "tactical innovation" recombine established tactical elements rarely if ever used together before (Wang & Soule, 2016). Similarly, successful popular music reflects trending tastes with differentiating novelty (Askin & Mauskapf, 2017).

Not all recombinant innovations are equal. Some are incremental, replacing one or more components with improved alternatives to exploit knowledge of proven combinations. Others are radical (Utterback, 1994) or architectural (Henderson & Clark, 1990), realized through exploration of new and uncertain combinations (Youn et al., 2015). From the complex systems perspective, innovators search over a high-dimensional "rugged landscape" of combinatorial possibilities and risk failure to make the greatest impact (Billinger et al., 2014; Fleming & Sorenson, 2004). Innovators hedge this potential for failure by building portfolios that balance more and less risk. For example, scientists often perform both innovative and traditional research to optimize rewards (Foster et al., 2015) and high-impact research often mixes novel and conventional components to balance novelty with familiarity (Uzzi et al., 2013). Modeling and measuring innovation as a combinatorial process, innovation studies have measured emergent innovation, and the process of making it, at scales ranging from macroscopic at the level of civilizations to microscopic at the level of individual social agents.

.

2.2 Innovation at societal scale

The innovation capacity of civilizations increases through destructive creation as disruptive and even traumatic shocks break down existing social, cultural and material structures, such as natural disasters and war. By breaking apart complex institutions, these shocks facilitate the process of creative recombination by dislodging parts of nature, culture and society previously unexposed and unavailable for recombination.

In the wake of war, social order is disrupted and resources and human capital are shuffled. Random connections between ideas and intellectuals proliferate, resulting in a rate of innovation uncommon in stable societies. Consider Paul Forman's historical analysis of Weimar culture and quantum mechanics. The loss of WW1 and the memory of horrors and chaos it unleashed on Europe, upended German society and deeply affected its intellectuals. Feeling that notions of 'causality' had inspired Germany's national identity and aggression in the war, German philosophers articulated an acausal nihilist worldview that broke down assumptions of rationality and determinism. Forman's detailed historical analysis demonstrated that physicists from the Weimar Republic not only read, but in some cases wrote philosophy in this vein. It was these and nearby physicists who imagined the possibility of quantum mechanics, an acausal physics, which would not have emerged at that time and place if hostility against science, rationality and materialism, including Newtonian causality, hadn't broken their connection with contemporary views of physical reality. Historians drew the 'Foreman thesis' from this demonstration, that culture furnishes at least part of the imagination underlying new scientific discoveries, and the milieux from which they become accepted. This hints at our more specific thesis that wars, natural disasters and other society-scale calamities break apart established certainties into fragments that may be recombined to form new ideas and institutions.

Society-scale conflicts and disasters not only supply, but demand new forms of innovation. Military and health-related funds and associated resources are unleashed during such periods, enabling transformative science, expertise and technology (Lee, 2016). For example, in the mid-20th Century, WWII catalyzed nuclear power and explosives, large-scale computation, radar, and a number of medical techniques (Gross & Sampat, 2020). For countries, desperate incentives for survival and competitive priority under novel circumstances call forth these technological breakthroughs (Hanlon & Walker Hanlon, 2015), which have wide ranging impacts once war is over. Recent work demonstrates that even economic crises, which result in reduced innovation investments among most firms, stimulate an increase of innovations among the most innovative firms exploring changed product and market environments (Archibugi et al., 2013).

2.3 Innovation within communities and networks

Beneath long-term cycles of societal conflict and catastrophe, innovation emerges more routinely from "creative conflict" (Stark, 2011) between groups and people with different perspectives and lived experiences. Such conflicts raise questions and break down the cognitive inevitability of existing structure within networks and teams.

Social network analysis provides a prominent perspective on innovations within communities. Classical network theories emphasize the importance of network positions in determining innovation capacity. Ron Burt demonstrated that opinions within business groups tend to be homogenous, but brokers that uniquely connect distant groups across "structural holes" in the network have a broader vision (Burt, 2004). These arise largely from the diversity and conflict of perspectives and preferences across the network. Brokered regions of the network break the unity of established understanding and illuminate opportunities for novel recombination across boundaries. This builds on prior work by Mark Granovetter regarding the "strength of weak ties". That work illustrated how weak or distant ties—connections with acquaintances or even strangers—provide differential access to novel information from outside the group (1973). Unique, high-value information that passes along weak ties bursts self-reinforcing communication bubbles by weakening the structure and hold of local information. In this way, cross-group communication can increase innovative performance by providing conflicting perspectives on the world (Wal et al., 2020).

The effect of network patterns on innovation also exists at the level of inter-organizational networks. Firms with higher network centrality tend to have better innovation performance not only because they have more control over information and resources (Bell, 2005), but because they have greater access to dissonant information viewable across otherwise disconnected individuals and organizations (Stark, 2011). The innovative power of conflicting and chaotic information requires changing sources. Stable, unchanged alliance networks reduce firm innovation, while adding those that span emerging structural holes renews innovative opportunity (Kumar & Zaheer, 2019). Rotating leadership between collaborating firms is another strategy that synthesizes complementary capabilities and yields fresh conflicts, opening the space for technological innovation (Davis & Eisenhardt, 2011). Network studies highlight the importance of occupying network positions that violate the reinforcing structure of local information and unleash the imagination and opportunity to innovate across social scales. Sustainable innovation results from the renewal of these positions as structural holes open and close across the landscape. Moreover, as in wars between countries, leading firms in winner-take-all markets have strong incentives to maintain technological innovation (Aghion et al., 2005). Likewise, intra-organizational conflict arising from competition, can sustain incentives to innovate (Kidder, 1981).

Innovation also occurs within teams proportion to the cognitive conflict they stage. Research on rap music (Lena & Pachucki, 2013), video games (De Vaan et al., 2015), political communication (Kreiss & Saffer, 2017), scientific discoveries and technical inventions have shown that radical innovations emerge when cognitively diverse teams combine previously disconnected ideas (Shi & Evans, 2019). Team size also influences the creative performance of these teams. Smaller, flatter teams tend to produce disruptive innovations by activating low-level cognitive conflict between team members holding diverse expertise (Wu et al., 2019; Xu et al., 2022).

Not only diverse teams, but teams differing from the majority and its established status quo are more likely to generate things perceived as innovative to that majority. Students from minority groups have been shown to innovate more (Hofstra et al., 2020), and minority dissent stimulates creative conflict and innovation within the team (C. K. De Dreu & West, 2001; Sutton, 2002). New artists are often associated with the emergence of new musical genres, more so than collaborations among established artists (de Laat, 2014). In the context of scientific and technical innovation, the greatest and most disruptive successes occur when "expeditions" of investigators from one area travel to another cognitively diverse region to solve their challenges in unprecedented and unfamiliar ways (Shi & Evans, 2019).

The cognitive conflict stirred by diversity may be sustained by supportive team, organizational and societal cultures (Cho, 2022; Hülsheger et al., 2009). When a social collaboration is stable, diverse ideas and perspectives may be exchanged efficiently, without endangering the collaboration itself. For example, research suggests that norms for avoiding sexist language reduces the tension and improves the creativity of mixed-sex groups (Goncalo et al., 2015). Equal opportunity to criticize and debate may encourage an atmosphere conducive to idea generation from all participants (De Dreu, 2006; Nemeth et al., 2004).

At this meso-level of networks and teams, *destructive creation* operates at the level of cognitive, cultural and social conflict to create the possibility for ongoing innovation (Jung & Lee, 2015).

2.4 Innovative conflict within individuals

Societal innovations almost always occur in social interaction (Duede & Evans, 2021) and come in multiples (Tria et al., 2014). As such, much social and behavior research focused on individual persons explores not how they innovate, but why. Economists and sociologists have focused on extrinsic rewards for innovation (Dosi, 1982) including recognition (Merton, 1973), promotion (Foster et al., 2015) and cash (Hamel, 2006). Several have demonstrated that shifts in rewards significantly affect the quantity and quality of innovative production (Foster et al., 2015; Hvide & Jones 2018; Myers et al., 2020; Yanadori & Cui, 2013). Different positions within the system may also yield differing incentives to innovate. For example, young firms benefit more from partnering with venture capitalists (VCs), who seek to benefit from disrupting the established industrial order, than with government agencies or corporate sponsors threatened by radical changes to the status quo (Pahnke et al., 2015). Beyond extrinsic motivations, philosophers and psychologists have emphasized intrinsic rewards that flow from desirable experiences associated with curiosity, challenge and exploration. These may motivate innovation independent from the extrinsic rewards mentioned above.[1]

Beyond this focus on personal motivations behind why people innovate, when psychologists have investigated the process of how they do so, they consistently find that conflict within the mind and body is associated with novel ideas, designs and action. These internal conflicts arise from the interaction of diverse skills, knowledge and experience within individuals, and from the shock of traumatic or challenging lived experiences.

Diverse experiences have been experimentally shown to enhance cognitive flexibility and the creation of novelty (Barbot, 2022; Ritter et al., 2012). Similarly, when individuals are embedded within complex environments and face conflicting demands, they may adopt paradoxical mental frames and become more likely to recognize and embrace contradictions, stimulating creativity (Miron-Spektor et al., 2011; Ritter et al., 2012). This patterns scales up: individuals with broader experience are capable of coordinating others across that breadth, yielding more collective creativity (Vries et al., 2014).

---

[1] In coining the term Creative Destruction, Sombart emphasizes the 'will to create' (Sombart, 1930) among entrepreneurs.

Other research characterizes creativity as arising from traumatic or deeply challenging personal experiences (Forgeard, 2013). Moverover, people who hold a conflict mindset—who are led to perceive others as their adversaries—have been shown to hold broader and more inclusive cognitive categories (De Dreu & Nijstad, 2008). Finally, jarring experiences with different cultures or facing divergent stereotypes, which challenge personal biases, can stimulate creativity (Crisp & Turner, 2011; Gocłowska et al., 2013). This ties in with the way diverse experiences create internal conflict and open the space for novel thinking.

3. Innovation Diffusion as Creative Destruction

Most prospective innovations fail (Castellion & Markham, 2013) from inadequacy, poor fitness within their environments, or when defeated by forces defending the status quo. Those that succeed elicit a cascade of diffusion among adopters. This stimulates a social process widely described as creative destruction. In creative destruction, old components are permanently abandoned or replaced, giving way to the new order. From a complex system perspective, the genesis of social, cultural and material novelty increases entropy or disorder in the system, but widespread diffusion establishes new organization, driving out extant diversity. This process of creative destruction follows the invention stage and is often conceptualized as "innovation" in a narrow sense. In practice, it often happens simultaneously with invention, and together they shape the innovation landscape.

3.1 Innovation Adoption

Innovation exposes adopters to excessive risks and costs, including adjustment and transaction expenses (Bigelow et al., 2019), loss of original advantages (Christensen et al., 2018), and uncertainty (Kim, 2020). Potential adopters lie at different points along a lazy S-shaped or sigmoidal curve of adoption growth (Bejan & Lorente, 2011; Shinohara & Okuda 2009; Shimogawa et al., 2012; Yuri & Federico, 2012). The shape of this curve implies that during early diffusion, the number of adopters is a relatively small proportion of all potentials, which accelerates until reaching the point of inflection, with infinite slope, beyond which it saturates toward an asymptote, given by the total number of eventual adopters. Numerous explanations exist for the S-shape, including one that emphasizes information flow among potential adopters, resulting in epidemic adoption with adoption rates proportional to numbers of "infected" and "uninfected" candidates (Shogren, 2013, p.65). Assuming that potential adopters are equally likely to interact, this process is modeled by the logistic differential equation, leading to a logit sigmoid function for cumulative adoptions, with the curve's inflection point representing a tipping point in the cascade of adoption.

An alternative "probit" explanation relies on heterogeneity among potential adopters. Everett Rogers classically partitioned adopters into categories, each with different values and risk preferences (Rogers, 2010 [1962]) arrayed at different points along the curve. Initial adopters may gain most from the intrinsic advantages of innovation quality or efficiency (Compagni et al., 2015; Catalini & Tucker, 2017), while later adopters shift with changes in costs to adoption by collective learning or the advantages of scale. Consider sports reporters joining Twitter to better spread their story (English, 2016) or politicians updating Facebook profiles to increase their competitive potential (Williams & Gulati, 2013), network effects that mix logit and probit dynamics with communication broadening access to the platform, which increases its value for adopters.

Institutional explanations transcend rationality by emphasizing how adopters conform to emergent rules in their environment and gain legitimacy through innovation adoption as "ceremony", even if counter-productive (Meyer & Rowan, 1977).[2] This has been observed in organizations adopting structures (DiMaggio & Powell, 1983), cities adopting civil service reforms (Tolbert & Zucker, 1983), and many instances of personal technological adoption. These findings resonate with extensive research on peer effects, broadly defined as family, neighborhood, and workplace influences that provide social opportunities and cultural pressures towards innovation adoption. People in organizations follow both advisors' and co-workers' attitudes when making innovation adoption decisions (Fu et al., 2020), just as they do in family and community (Correa et al., 2015). By contrast, social and cultural disconnection reduces adoption convergence, with lower firm connectivity within cities associated with increased knowledge diversity (Esposito & Rigby, 2018) and disconnected cities open to more ambiguous jazz recordings than central hubs (Phillips, 2011).

---

[2] We note that Roger's innovation adopters, while facing "a high degree of uncertainty about an innovation at the time he or she adopts", are also influenced by information and social factors (Rogers 2010 [1962]).

In all of these explanations, local social and cultural structures protect and channel potential innovations to widespread adoption against conservative social and cultural forces that defend the status quo and resist replacement. Social networks represent a critical substrate conveying information access and unleashing local pressures for conformity (Podolny, 2001). Social networks have been used to explain a wide range of diffusions, from the poison pill practice among firms (Davis, 1990) to ISO9000 quality certificates among nations (Guler et al., 2002). The combination of network effects and attributional or cultural similarity unleashes accelerated diffusion. This occurs through "homophily", whereby individuals are attracted to others like themselves, a process demonstrated in experiments tracing the spread of personal health behaviors (Centola, 2011). Because of network and attributional clustering, even though an innovation may be rare in the system, it can represent a local majority, unleashing enticements and pressures for conformity among those nearby. The development of computational tools has made it possible to better infer and simulate the diffusion process among a complex field of potential adopters. Recent examples include policy diffusion among U.S. states with the NetInf algorithm (Desmarais et al., 2015), the diffusion of Martin Luther's revolutionary religious ideas with epidemiological modeling (Becker et al., 2020), or the complex diffusion of new norms and style with agent-based modeling (Centola & Macy, 2007; Centola et al., 2005; Manzo et al., 2018).

Spatial proximity reflects another type of local structure that protects and propagates innovation (Boschma & Frenken, 2010; Phene et al., 2006; Zhang et al., 2009). Proximity accelerates communication and interaction, providing the basis for common experience, mutual understanding (Gertler, 2003), and local pressures for conformity. Spatial proximity accelerates communication and interaction (Storper & Venables, 2004). For example, spatial proximity corresponds to the spread of protest tactics among colleges and universities with similar features (Soule, 1997) and political behaviors among geographically and politically proximate counties (Xu & Tian, 2020).

Adoption may also reflect conformity to resonating fashions and values. For example, research has demonstrated the appeal to professionalism in reporters' adoption of "fact-checking" (Graves et al., 2016) and the acceptance of environmentalism among rich populations with post-material values (Pampel & Hunter, 2012). Values may also be imputed, as in the largely male practice of bird photography in the early 20th Century (Song, 2018). Fashions also shape the path of diffusion (Abrahamson, 1996). When management innovations become widely demonstrated as useful[3], they diffuse widely and become fashionable (Birkinshaw et al., 2008), creating widespread appreciation for an innovation that transcends utility and constructs cascades of emulation. When the average quality of adopters falls, as from "ceremonial" or pressured adoption, consultants jump to newer innovations and the fashion cycle continues (Strang et al., 2014). This dynamic is also reflected in the rise and fall of scientific techniques and approaches (Bort & Kieser, 2011).

Beyond factors mentioned above, social structures such as political balance (Parinandi, 2020) and innovator migration (Fackler et al., 2020; Ferrucci, 2020), cultural structures like legal intellectual property rights (Cockburn et al., 2016), and material structures like platforms and virtual spaces (Claussen & Halbinger, 2021; Halbinger, 2018) also impact innovation diffusion.

3.2 Innovation Evolution and Complex Interaction

---

[3] The converse is also demonstrated—weakly mobilized innovation supply may be creditable for diffusion failure (Ax and Ax 2021).

Frameworks of innovation adoption and diffusion have been criticized in social and cultural accounts as insufficiently reflecting changes in the focal innovation as it shifts context. Political science studies, however, have broadly explored how borrowed policies shift with policy context. Social learning changes the nature of an innovation: if states have already widely adopted a policy, new potential adopters learn from previous experience and focus on practical over normative dimensions when considering it afresh (Gilardi et al., 2021). Once adopted, policies also undergo postadoption reinvention and amendment (Carley et al., 2017).

Research beyond the social sciences identifies material properties and interactions that shape innovation diffusion and evolution. Innovation complexity has been shown to negatively correlate with diffusion speed (Plsek, 2003). Complex interrelationships between innovations also shape and constrain the adoption process. At the extremes, innovations may mutually support one another to accelerate symbiotic diffusion (R. J. Thomas & Wind, 2013), or compete to exclude one another with the winner dominating.

Symbiosis or competition may occur between social, cultural, and material innovations. Consider blockchain technologies that seek—within only partial success—to replace social or institutional trust in a bank, country, or legal system with a distributed network of duplicated, immutable ledgers (Hawlitschek et al., 2018; Werbach 2018a, 2018b). Innovation unleashes complex cascades of adoption and evolution. Social innovation can lead to technological innovations (Mol & Birkinshaw, 2013) and technological evolution impacts the social system (Arthur, 2009). Nevertheless, social scientific studies of innovation prefer motivational, social and cultural causes. This has led to an under-appreciation of the complex dynamics involving social, cultural *and* material forces more widely considered in complex systems and science studies. At the extreme, the actor-network tradition grants material forces influence and even imagines agency as "actants" in an interacting system (Callon, 1986; Callon, Arie et al., 1986; Latour, 1987).

Diffusion success and failure conditions the ultimate impact that innovations have on their inventors, adopters, and society as a whole.

4. Innovation Outcomes

4.1. Progress, Failure and the Invisible Hand of Innovation

Innovation has been closely related to the concept of social and natural progress. In economics, this idea finds expression in endogenous growth theory, where successful technological innovation contributes to growth, leading innovation investments to follow expectations in a recursive cycle (Jones, 1995; Romer, 1983; Segerstrom, 1998). Innovation is also attributed a critical strategy for individual, scientific, organizational and national competitive leadership and differentiation (Doms et al., 1995; Foster et al., 2015; Porter, 2011; Porter, 1985; Roberts, 1999; Sinha & Noble, 2008; Wojan et al., 2018). These advantages, are mediated by cultural acceptance, and are often discounted for minority or disadvantaged innovators (Hofstra et al., 2020).

While innovation may bring reward, it also risks failure. This manifests in higher outcome variance in technology (Fleming, 2001), failure to publish or receive citations in science (Foster et al., 2015; Uzzi et al., 2013) and speculative portfolio risk in financial markets (Simsek, 2013). It also leads to cognitive and community resistance. When a product is novel and does not clearly belong to a current category, critics and consumers cannot easily make sense of its identity and make lower evaluations from confusion (Goldberg et al., 2016; Hsu et al., 2009; Leung & Sharkey, 2014; Pontikes, 2012; Shi et al., 2018; Zuckerman, 1997). But institutionalized boundaries that highlight identity ambiguity feed a new cycle of innovation genesis. In software markets, Pontikes contrasted the role of market "takers" who consume goods in avoidance of ambiguity with market "makers" who redefine market structures drawing on ambiguity as opportunity (2012). As we discuss below, established institutions heighten cognitive awareness of their violation and sow the seeds of their own disruption.

4.2 Societal Innovation and Paradox

If innovation genesis and adoption characterize societal innovation, how do contemporary economies, cultures, or scientific institutions depress or amplify it? U.S. and OECD economies have seen a marked decline in business dynamism. Market concentration has increased, oligopolies have emerged, diversity and labor productivity has dropped, knowledge diffusion has stalled, and quality-improving innovations from incumbent firms increasingly outweigh destructive innovations by entrants (Akcigit & Ates, 2019; Akcigit & Ates, 2021; Akcigit et al., 2021; Dowd, 2004; Garcia-Macia et al., 2016). While cultural innovations like the emergence of rap have changed the popular music scene, other cultural industries, from Broadway to Orchestras have become conservative, especially for incumbents, partly to satisfy conservative tastes held by elite consumers (Kremp, 2010; Phillips & Owens, 2004; Thomas, 2019). Science has slowed in innovation as well. The number of scientists and papers has increased dramatically over the past few decades, but individual scientist productivity has declined (Bloom et al., 2017; Jones, 2009; Thomas, 2019). Paradoxically, increases in research quantity have slowed scientific progress by making it harder for new ideas to gain widespread attention (Chu & Evans, 2021). Citations flow to well-known researchers, making it more difficult for outsiders to challenge field leaders (Zivin et al., 2019). Furthermore, fast-paced research gives a competitive edge to specialists, while research spanning disciplines is slower and involves greater risk (Leahey et al., 2017; Teodoridis et al., 2019). These trends highlight a driving force underlying the paradox of innovation science. Innovative success enriches and empowers once-innovative individuals and institutions to defend against new waves of innovation, which forces novel sources of change to emerge from new quarters of science, technology and society.

Marx defined the innovation paradox by arguing technological advances in capitalist society increase the rate of relative surplus-value, narrowing the accumulation of capital to the winners of industrial competition, enlarging the wealth gap and ultimately incentivizing the destruction of capitalism from the inside (Marx, 2013). Recent studies have provided empirical quantitative support for rising inequality driven by capitalist competition (Piketty, 2014; Piketty & Zucman, 2014).

4.3 Supportive Mechanisms for Societal Innovation

Government policies may play a role in shaping the trajectory of societal innovation. As in teams, government policies that support integration and broad collaboration encourage sustained innovation. Consider how policies that improve racial integration (Samila & Sorenson, 2017) encourage broader, more diverse participation, which increases the quantity and quality of creative activities and products for society. Supportive policies can also facilitate knowledge diffusion (Bloom et al. 2017; Jandhyala & Phene, 2015; Thomas, 2019), accelerate innovation, and broaden its relevance for more of society (Koning et al., 2021). Such policies activate and draw upon the benefits of sustained diversity through contained cognitive conflict.[4]

With mass digitalization and the development of the internet, new social forms of innovation have arisen like crowdsourcing, which recruits a population of crowd-workers on a platform through open call and assigns them roles in a task. As crowdsourcing takes greater advantage of distributed knowledge, it opens innovative opportunities (Schenk et al., 2009). Crowd sourcing is one emerging method of "open innovation" whereby firms and government agencies (e.g., U.S. NASA, DARPA programs) broadly source ideas within and beyond their members to advance innovative performance (Chesbrough et al., 2008). The internet also facilitates distanced collaborations, but our own work has demonstrated that the most surprising and influential scientific and scholarly ideas arise from in-person contacts at diverse institutions like universities (Duede et al., 2021), reinforcing concern that the innovative social form of online crowds might not themselves be engines of auxiliary innovation (Brucks & Levav, 2022). Moreover, systemwide collaborations accelerated by contemporary internet connectivity unleash innovative combinations in the short-term, but paradoxically lead to a collapse in cultural diversity on which recombinant innovation relies (Li et al., 2020).[5]

Additional innovative forms are emerging from the introduction of artificial intelligence (AI) agents in society. Exploration of the potential for AI collaboration within scientific discovery (Gil et al., 2014), business (McAfee & Brynjolfsson, 2017) and social life is accelerating (Liu, 2021), though still underrealized (Barro & Davenport,

---

[4] Legal policies, including patent law and other intellectual property protection, have also exerted a profound motivational impact on social innovation outputs (Moser, 2005; Moser and Voena, 2012).

[5] Recent accounts of technological progress based on digitalization and AI document the replacement of workers with enlarging gaps in wealth, income and mobility. Despite unleashing innovation, they also contribute to a widening of class(ic) inequalities (Brynjolfsson and McAfee 2014).

2019). It is notable that rapid advances in deep neural networks, the workhorses of much modern AI themselves draw not only on computer science, but conflicting ideas and technical designs from statistics, cognitive and neuroscience, mathematics, physics, evolutionary algorithms, behavioral psychology, and many other fields in generating improvements (Domingos, 2015; LeCun et al., 2015).

5. The Future Science of Innovation

In this review essay, we have sought to trace the theoretical line of innovation studies across the social sciences. Social science theory and methods deeply shape how we understand innovation. As we have shown above, the intensive structure and regulation of contemporary society, culture and technology lead to a cascading genesis of innovations in one domain when established auxiliary structures and defensive forces break down. Resulting conflict and chaos unleash the imagination, invention and adoption of innovations, a process we frame here as *destructive creation*. This cyclically precedes and follows the companion process of *creative destruction*, wherein emergent innovations that are nurtured, protected and disseminated by local knots in social, cultural and technical structure, spread and render those that have gone before irrelevant.

As the paradox of innovation science suggests shifting sources of innovation in the world, innovation in understanding about innovation will likely emerge from beyond the social sciences. Complex relationships between innovation and its outcomes are not linear, and remain underexplored in social science. Economics has focused on rational human action driving accumulated progress, psychology on personal motivation and proclivity, and sociology on social and cultural institutions, but all dismiss the primacy of material forces as deterministic, despite their widespread appreciation in earlier eras of social investigation (Ogburn, 1947), the natural sciences from biology to physics (Diamond, 2013; Le Couteur, 2003; Mukherjee, 2010), and science and speculative fiction (Heinlein, 1947), from hard science fiction to apocalyptic and climate fiction. These have consistently imagined social, cultural and material consequences of emerging technologies and their externalities.

The complex role of material forces has also been imagined but rarely measured by formal theories from complex systems, drawn from physics and biology (Simon, 1991). They have further been articulated in rich cases and grounded theories from science and technology studies (Bijker & Law, 1994; Bijker, 1997; Callon, 1986; Pinch & Bijker, 1984; Latour, 1987), but rarely assessed at scale. In Figure 1 below, the process of destructive creation moves from upper right to lower left, as creative destruction moves in reverse. If, as in the Figure 1, we anchor innovation research by its focus on social, cultural and material forces, we will find that there is an empirical hole in the center of the field. We argue that this hole represents an innovation opportunity and challenge that could be supplied by big innovation data and emerging computational approaches.

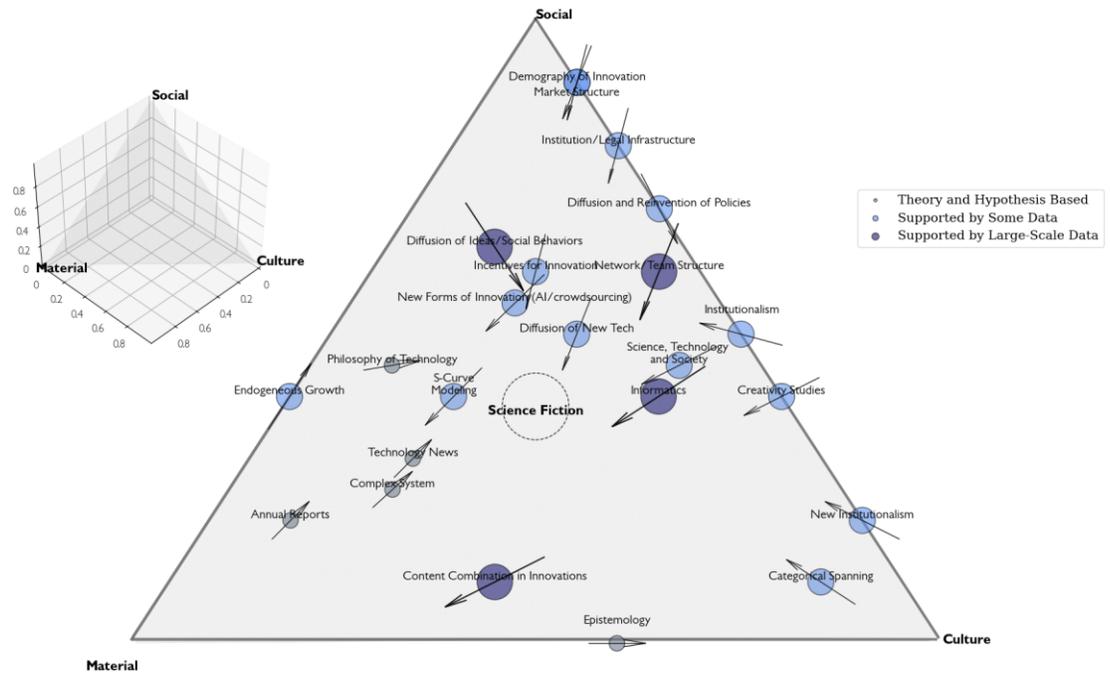

**Figure.** Mapping of research literature related to innovation science. Each literature represents both a point and a vector within the social, material and cultural factors anchoring this two-dimensional simplex (illustrated in the upper left). Points represent relative attention to the factors, size represents magnitude of the extant literature, direction represents dominant causal orientation from one factor to another. Darkness loosely represents stronger empirical support within the literature (e.g., more data and quantitative analysis); lightness represents more theoretical, small-scale or selective empirical support.

Empirical innovation research places unique demands for data. Data on a modal pattern requires at most half as much data as research on a dynamic or spatial trend if anchored by only two points (Evans & Aceves, 2016). Claims about emergent novelty or innovation require a much more complete representation of the prior period or domain, risking the characterization of something as new that lay in the unexamined tail of the frequency distribution. This challenge is faced by "origins" research in all fields—from the origins of space, matter and life to technology, language and social organization—where new generations of research succeed trivially by pushing origins back further in time or across space.

Widespread digitization and the expansion of inexpensive sensors—from social media to $CO_2$ trackers—provide theoretical and methodological tools to trace patterns of activity and behavior at scales previously unimaginable. For example, large-scale data from online platforms have made it possible to model innovation evolution during diffusion, as in the analysis of Facebook memes. Such analysis demonstrates how meme mutations can be well modelled by the Yule process, a statistical model widely used in modeling genetic phenomena (Adamic et al., 2016). Machine learning approaches are rising to meet the challenge of large-scale digital data, enabling dimension reduction, multi-modal embedding and massive alignment across data types. We and others have begun to use these data and computational approaches to directly integrate material, social and cultural innovations (Shi et al., 2015; Foster et al., 2021), and to model complex processes of innovation that combine these (Sourati & Evans, 2021; Shi & Evans, 2019), building on prior science studies inspiration (Callon, Arie et al., 1986). Of course, even these opportunities hide conceptual costs and raise boundaries between the empirical examination of innovative domains for which data is available and for which it is not.

Our review has explored and identified frontiers of research attention in the examination of innovation (1) genesis, (2) diffusion, and (3) consequences for the world. This reveals how social scientists naturally focus far more on the social and cultural contexts of innovation than its contents; framing and seeking to identify them as causal. More specifically, our review reveals the contemporary bias in viewing social and cultural disorder as the primary driver of material innovation, rather than the converse. We argue that innovation studies could become more powerful, descriptive and predictive, if and when they saw social, material and cultural forces as co-constitutive and co-evolving, acknowledging the influence of technological innovations on society and culture as suggested by the fields of science studies and science fiction. We highlight computational tools that open new possibilities for analysis of novel content and context in interaction, and show how this brings us empirically toward the full range of ways in which humanity has imagined connecting the social, cultural and material poles of innovation, bridging the social study of innovation and innovation policy with innovation practices and advance.

Appendix A. Literature Searching Process and Criteria for Inclusion

This review focuses on both classics and recent publications about innovation in the social sciences. To achieve this goal, we aimed to identify the most influential works in the academic corpus. Here are details of our literature search process.

For the first step, we chose the most highly cited journals in the social sciences, as shown in Table A1. From the publications in these journals, we kept all papers with the keywords "innovation", "invention" and "creation" in their titles, keywords, or abstracts, and set the time frame from 2011 to 2022. We constrained the search results to be "articles", and this process left us with a sample of 1679 articles.

Table A1. Representative Journals as the Starting Point for Literature Searching

| Discipline | Journal Names |
| --- | --- |
| Sociology | *American Sociological Review (ASR), American Journal of Sociology (AJS), Social Forces, Social Forum* |
| Management | *Administrative Science Quarterly (ASQ), Academy of Management Journal (AMJ), Strategic Management Journal (SMJ), Management Science, Organization Science* |
| Economics | *Econometrica. American Economic Review (AER)* |
| Communication | *New Media & Society, Journal of Communication* |
| Politics | *American Journal of Political Science (AJPS), American Political Science Review (APSR)* |
| Psychology | *Annual Review of Psychology, Psychological Science* |
| Other Interdisciplinary Journals | *Science, Nature, Proceedings of the National Academy of Sciences of the United States of America (PNAS), etc.* |

For the second step, we (1) ruled out some papers not directly related to the innovation process, and (2) kept a subsample of the papers with similar topics. One example for the first category is the 2021 publication in *Organization Science*, "When Do Firms Trade Patents?" This paper is about the market for innovation, but not directly related to innovation itself, so we deleted it from our sample. For the second category, one example is the collection `of studies about team composition and innovation results (N≈40). Although these papers ha`d their unique conclusions and contributions, they shared a similar perspective and methods. For these clusters of studies, we only kept one or two studies for each family, based on their influence (measured with citations) and relevance. When sociology papers and studies from other disciplines appeared together in one cluster, we gave preference to sociology papers. The second step left us with approximately 120 papers.

For the final step, we made selective additions, including (1) classic studies that provide a theoretical basis for later work—an example is Rogers (2010, [1962]) that laid the foundation for innovation diffusion theory; and (2) newer works repeatedly mentioned and cited by other authors, including the book of Padgett and Powell (2012), which that applied the complexity perspective to social and organizational innovation. This step left us with just over 190 studies cited in this paper.